# Low Frequency Carrier Kinetics in Perovskite Solar Cells

*Vinod K. Sangwan,[#] Menghua Zhu,[1,2,#] Sarah Clark,[1,#] Kyle A. Luck,[1] Tobin J. Marks,[1,3] Mercouri G. Kanatzidis,[3] and Mark C. Hersam[1,3,4,]**

[1]Department of Materials Science and Engineering, Northwestern University, Evanston, IL 60208, USA
[2]Wuhan National Laboratory for Optoelectronics (WNLO), Huazhong University of Science and Technology (HUST), Wuhan 430074, China
[3]Department of Chemistry, Northwestern University, Evanston, IL 60208, USA
[4]Department of Electrical Engineering and Computer Science, Northwestern University, Evanston, IL 60208, USA

[#]These authors contributed equally

*E-mail: m-hersam@northwestern.edu

**Abstract**

Hybrid organic-inorganic halide perovskite solar cells have emerged as leading candidates for third-generation photovoltaic technology. Despite the rapid improvement in power conversion efficiency (PCE) for perovskite solar cells in recent years, the low frequency carrier kinetics that underlie practical roadblocks such as hysteresis and degradation remain relatively poorly understood. In an effort to bridge this knowledge gap, we perform here correlated low frequency noise (LFN) and impedance spectroscopy (IS) characterization that elucidates carrier kinetics in operating perovskite solar cells. Specifically, we focus on planar cell geometries with a $SnO_2$ electron transport layer and two different hole transport layers – namely, poly(triarylamine) (PTAA) and Spiro-OMeTAD. PTAA and Sprio-OMeTAD cells with moderate PCEs of 5 – 12% possess a Lorentzian feature at ~200 Hz in LFN measurements that corresponds to a crossover from electrode to dielectric polarization. In comparison, Spiro-OMeTAD cells with high PCEs (>15%) show four orders of magnitude lower LFN amplitude and are accompanied by a cyclostationary process. Through a systematic study of more than a dozen solar cells, we establish a correlation with noise amplitude, power conversion efficiency, and fill factor. Overall, this work establishes correlated LFN and IS as an effective methodology for quantifying low frequency carrier kinetics in perovskite solar cells, thereby providing new physical insights that can rationally guide ongoing efforts to improve device performance, reproducibility, and stability.

**Introduction**

Hybrid organic-inorganic halide perovskite solar cells (PSCs) have quickly risen to the forefront of photovoltaic research with certified power conversion efficiencies (PCEs) exceeding 23%.[1-5] Prominent photovoltaic materials in this class possess a tetragonal crystal structure of the form $ABX_3$ (A = $CH_3NH_3^+$; B = $Pb^{2+}$; and X = $Cl^-$, $Br^-$, $I^-$).[4] High PCEs in these solar cells originate from a fortuitous combination of desirable properties such as high panchromatic absorption,[6] sharp absorption edge,[7] long carrier lifetime and diffusion length,[8] low exciton binding energy,[9] intrinsic defects forming only shallow states,[10] and efficient charge collection at interfaces.[2-4, 6] Low cost processability, either in mesoscopic or planar architectures, further enhance the appeal of PSCs compared to competing organic photovoltaics or dye-sensitized solar cells.[3-4] In addition, not only is the 1.55 eV bandgap close to the ideal Shockley-Queisser limit, but it is also tunable with mixed cation or mixed halide composition, thus enabling tandem and multijunction cells.[5, 11-12]

Despite these many attributes, widespread deployment of PSCs is hindered by concerns surrounding light-induced ion migration, hysteresis, and stability.[13-14] Although encapsulation strategies have been shown to improve device stability by mitigating photo-oxidation and minimizing moisture intrusion, PSCs remain susceptible to intrinsic degradation mechanisms involving phase changes, ion migration, and chemical reactions.[13] Therefore, a fundamental understanding of these processes is critical to the development of rational approaches for improving stability under working conditions. Towards this end, several methods have been employed to probe correlated electron and ion kinetics in PSCs including impedance spectroscopy (IS),[6, 14-15] open circuit voltage decay measurements,[16] and transient absorption in addition to photoluminescence and photocurrent spectroscopy.[8] Another experimental approach that is commonly used to quantify carrier kinetics in conventional semiconductors is low frequency noise (LFN) characterization.[17] Dark noise also plays a critical role in other optoelectronic applications of hybrid organic-inorganic perovskites such as light-emitting devices,[18] photodetectors,[19] and lasers.[4] Early efforts to apply LFN characterization to perovskite materials and devices include observed increases in LFN at the tetragonal to orthorhombic phase transition in perovskites at 150 K,[20-21] probing trap energy levels in native perovskites films with LFN,[22] and correlating LFN amplitude with PCE metrics in PSCs with different interfacial layers.[23] While these preliminary reports have yielded some important insights,[20-23] LFN remains an underutilized tool to probe the electronic and ionic processes that are responsible for hysteresis and degradation in PSCs.

Here, we report a correlated study of LFN and IS in planar methylammonium lead iodide ($CH_3NH_3PbI_3$) perovskite solar cells. Although initial PSCs were optimized using a mesoporous scaffold of the electron transport layer (ETL), excellent electron transport within the absorbing perovskite layer also yields high PCE in a planar geometry. For example, PSCs using planar $SnO_2$ and mesoporous $TiO_2$ achieve comparable open circuit voltage ($V_{OC}$) with PCE of 20%.[24-25] The planar geometry also offers a well-defined platform to study interfacial processes that are likely to be revealed by LFN and IS characterization. For the ETL, $SnO_2$ is among the most promising materials due to its low temperature (< 200 °C) growth and thus higher suitability for flexible substrates compared to mesoscopic $TiO_2$ that requires high temperature (~500 °C) sintering.[26] Furthermore, the ETLs utilized here exhibit near-ideal transparency, charge transfer, and achieve high carrier collection efficiency. On the other hand, the $V_{OC}$ and ultimately PCE of PSCs could be potentially improved by understanding limitations to charge transfer, carrier collection, and $V_{OC}$ resulting from the hole transport layer (HTL).[27] Therefore, in this study, we compare two HTLs – poly(triarylamine) (PTAA) and 2′-7,7′-tetrakis(N,N-di-p-methoxyphenylamine)-9,9-spirobifluorene (Spiro-OMeTAD) – that possess tradeoffs in PCE and device stability. In particular, Spiro-OMeTAD cells have shown the highest PCE albeit with limited device stability. In contrast, PTAA cells, possess reduced hysteresis at the expense of compromised PCE.[1, 27-28]

**Experimental Section**

We fabricated two classes of planar cells – $FTO/SnO_2/CH_3NH_3PbI_3$/Spiro-OMe TAD/Au and $FTO/SnO_2/CH_3NH_3PbI_3$/PTAA/Au – by spin coating followed by annealing and thermally evaporating gold contacts (see Experimental Methods for details).[29] All processing steps were carried out in a nitrogen glove box except the deposition of Spiro-OMeTAD, which was performed in a dry box. Fig. 1a shows a top-view scanning electron microscopy (SEM) image of a polycrystalline perovskite film with typical grain size of 100 – 300 nm, while Figs. 1b and 1c provide side-view SEM images of completed Spiro-OMeTAD and PTAA cells. The average thicknesses of the ETL and perovskite layers are approximately 60 nm and 400 nm, respectively, in all cells. For the HTLs, the average thicknesses of the PTAA and Spiro-OMeTAD layers are 50 nm and 200 nm, respectively. The apparent thickness variations in the perovskite layer originate from the roughness of the underlying FTO substrate. X-ray diffraction of the perovskite film further confirms its tetragonal phase with minimal residual $PbI_2$ (Fig. 1d).[30]

Dark current-voltage (I-V) characteristics obtained at a bias sweep rate of 0.014 V/s in vacuum (pressure = $2 \times 10^{-5}$ Torr) show negligible hysteresis for PTAA cells and significant hysteresis for Spiro-OMeTAD cells (Fig. 1e), in agreement with literature precedent.[1, 28] Solar cell measurements under AM 1.5G light in ambient (Fig. 1f) show comparable hysteresis and short-circuit current density ($J_{SC}$ ~21.8 mA/cm$^2$) for the two cells. A total of 26 PTAA cells and 14 Spiro-OMeTAD cells were fabricated and used for PCE characterization (see the histogram in Supporting Fig. S1). A subset of 8 PTAA cells and 8 Spiro-OMeTAD cells were used for 1/f noise and impedance spectroscopy. In the following discussion, we focus on three representative cells: a PTAA cell and two Spiro cells of varying performance. The average values of PCE and FF for the PTAA cell are 10.8% and 50.3%, respectively. The Spiro-OMeTAD devices, which showed ~100 mV higher $V_{OC}$ than the PTAA cells, were categorized into two groups according to performance: (1) cells with high PCE (average ~ 15.4%) are labeled 'Spiro 1'; (2) cells with moderate PCE (average ~ 11.9%) are labeled 'Spiro 2' (see Supporting Fig. S2 for representative I-V curves of Spiro 2 cells, and Supporting Table S1 for a summary of performance metrics for the representative PTAA, Spiro 1, and Spiro 2 PSCs used in the detailed analysis). Reduced PCE in Spiro 2 cells is likely caused by imperfections incorporated during processing.

We conduct all LFN and IS measurements in the dark because optical irradiation not only enhances ion-induced polarization[31] but can also change the underlying structure of the perovskite.[32] Low frequency current fluctuations are measured by biasing the anode electrode (Au) at voltages lower than the built-in voltage ($V_{bi}$) that results in a non-equilibrium state. The current noise power spectrum density PSD ($S_I$) of a PTAA cell as function of frequency (f) shows a $1/f^\beta$ behavior with β in the range of 1.06 – 1.26 for f < 100 Hz (Fig. 2a). An ideal value of β = 1 suggests a flat density of states for fluctuations over a broad energy range. The current dependence of PSD, $S_I \sim I^{1.5}$, deviates from the ideal $I^2$ behavior of equilibrium phenomena (Fig. 2b),[33] which suggests that resistance fluctuations are not the sole origin of the current fluctuation and thus some of the measured 1/f noise is driven by the applied current. This type of behavior is commonly observed in electrochemical and organic photovoltaic devices, and likely occurs in PSCs due to ion migration. Noise PSD measured for a larger frequency range (up to 6 kHz) reveals a transition from 1/f behavior to $1/f^3$ behavior at a characteristic frequency ($f_C$) of 180 Hz (Fig. 2c). In this case, the noise spectrum is fit to:

$$S_I = \frac{A \cdot I^{1.5}}{f^\beta \left(1+\left(f/f_C\right)^2\right)} \quad (1)$$

where A is noise amplitude. Such a Lorentzian feature results from the dominance of a single type of fluctuation with a characteristics frequency $f_C$. Similar Lorentzian behavior has been observed in polymer/fullerene cells where $f_C$ is related to photocarrier lifetime.[34] Other LFN studies on PSCs showed incidental Lorentzian bumps buried under dominant 1/f noise features occurring around $10^2$ Hz with limited understanding of their origin.[20, 22-23] In another LFN study on ITO/PEDOT:PSS/CH$_3$NH$_3$PbI$_3$/PCBM/BCP/Ag cells, Lorentzian features were found to occur at higher frequencies of 15 kHz,[21] and were explained by sub-bandgap states in the bulk rather than interfacial ions.

In an effort to gain further insight into the origin of the Lorentzian feature, we conducted IS under identical conditions (Fig. 2d). The Cole-Cole plot of the PTAA cell agrees well with the equivalent circuit model shown in the inset of Fig. 2e, where $R_S$, $R_1$, and $C_1$ account for the high-frequency behavior and represent the series resistance, recombination resistance, and geometrical capacitance, respectively, where the latter two elements are bulk effects. Two additional pairs of elements, $R_2/C_2$ and $R_3/C_3$, are needed for good fits at low and intermediate frequencies, and account for interfacial impedances. While the exact origin of these elements remains a topic of debate, they likely originate from charge accumulation, charge transfer resistance, and/or additional interfacial electronic states.[6, 35] At high biases near $V_{OC}$ (~1 V), an inductor element, L, is used in place of $C_2$ to fit the small region of negative capacitance observed at low frequency. Furthermore, for biases between 0 V and 0.7 V, a constant phase element, with an exponent value of $0.8 \leq n \leq 0.93$, is used in place of $C_3$ for improved fits.[36] Model fits over the full frequency range (1 Hz – 1 MHz) are provided in Supporting Fig. S3, and the extracted parameters for each equivalent circuit element are delineated in Supporting Fig. S6.

Unlike organic photovoltaics, the chemical capacitance in PSCs is much smaller compared to other capacitances due to the significantly lower density of defects in the perovskite layer, thus allowing chemical capacitance to be omitted from PSC impedance analysis.[37-38] On the other hand, at low frequencies, the accumulation capacitance $C_S$ (approximately $C_2$ at low biases) can be up to two orders of magnitude larger than the high-frequency geometrical capacitance ($C_1$) of the perovskite layer (Fig. 2e). This excess capacitance is different from the capacitance caused by roughness of the electrodes.[39] Instead, this phenomenon is unique to PSCs with a physical origin that is akin to electrode polarization in liquid or solid-state electrolytes.[40] In particular, since hybrid organic-inorganic perovskites are mixed ionic-electronic conductors with a soft lattice framework, mobile ions accumulate near interfaces under applied biases (Fig. 2f).[28, 31, 41-42] It should be noted that the concept of electrode polarization is distinct from chemical capacitance, which has been shown to have limited utility for the p-i-n diode architecture of perovskite cells. However, accumulated ions only respond to AC signals at low frequencies, resulting in dielectric polarization being recovered at $f > f_S$, where $f_S$ is the transition frequency below which electrode polarization begins to dominate. Here, $f_S$ is typically on the order of a few hundred Hz in planar cells in the dark but has been shown to reach up to a few kHz under illumination.[41] In Fig. 2e, the capacitance-frequency (C-f) plot for the PTAA cell shows $f_S$ ~200 Hz, which agrees well with the Lorentzian $f_C$, and is the first indication that accumulated ions near the $SnO_2$ layer comprise the dominant fluctuators in the LFN spectra (Fig. 2f).

The LFN behavior of Spiro-OMeTAD cells further supports the proposed origin of current fluctuations. Noise PSD for a Spiro 1 cell (PCE > 15%) follows a $1/f^\beta$ behavior with $\beta = 0.83 - 1.21$ for $f < 10$ Hz at different biases (Fig. 3a). The current normalized PSD ($S_I/I^2$) is ~$10^4$ times smaller than PTAA cells. However, the PSD increases significantly with frequency, ultimately forming a local maximum at $f_{max}$ ~ 200 Hz. This type of cyclostationary process was observed previously in organic photovoltaic cells, and was simulated using kinetic Monte Carlo simulations.[43-44] Specifically, space charge instability creates a superposition of time-domain periodic potentials, resulting in a cyclostationary peak in the noise spectrum at a frequency corresponding to the characteristic wait time for trapping/de-trapping events.[43] For example, in field-effect transistors, an external periodic potential is applied to reduce 1/f noise amplitude with the resulting noise spectra showing cyclostationary peaks at higher frequencies.[45] Here, an intrinsic cyclostationary process could contribute to the reduced $S_I/I^2$ in Spiro 1 cells as compared to PTAA cells in the 1/f regime. While deep traps were implicated for cyclostationary processes in organic photovoltaics, they are unlikely to be the main contributor here because $f_{max}$ is two orders of magnitude below the frequency range previously associated with bulk trap states in PSCs.[21] In addition, similar to $f_C$ in PTAA cells, $f_{max}$ is strongly correlated with $f_S$ (Fig 2e). Thus, mobile ions near interfaces are more likely to play the dominant role in cyclostationary processes in Spiro 1.

Spiro 2 cells possess a minor cyclostationary peak only at small biases below 0.2 V (Fig. 3b). At low frequency, Spiro 2 cells show $1/f^\beta$ behavior with $\beta = 1.13 - 1.25$. In addition, the $S_I/I^2$ levels are approximately three orders of magnitude higher than Spiro 1 cells, but are still 5-10 times smaller than PTAA cells (Fig. 3b). At high biases, the cyclostationary peak converts into a Lorentzian feature at the same frequency $f_C \sim f_{max} \sim 200$ Hz as the PTAA cells, suggesting a common origin. The current dependence of noise PSD for Spiro 1 and Spiro 2 cells exhibit $\sim I^{1.22}$ and $\sim I^{1.79}$ power law behavior, respectively (Fig. 3c). A power exponent closer to 2 suggests that Spiro 2 cells have attained a greater equilibrium than Spiro 1 cells, which could be explained by trapping of mobile ions near the interface. Similarly, the dark I-V characteristics of Spiro 2 cells show reduced capacitive hysteresis compared to Spiro 1 cells (Supporting Fig. S2) and what appears to be greater non-capacitive hysteresis, which is characterized by positive values for the reverse scan in the voltage range of 0.6 V to 1.0 V. Lower capacitive hysteresis has previously been correlated with lower PCE for Spiro-OMeTAD devices where ions were immobilized by light soaking.[46] Non-capacitive hysteretic currents are understood to be distinct from the easily reversible charge accumulation at the ETL/perovskite interface, and have been linked to internal rearrangement of the inorganic scaffold, interfacial modifications, and degradation at the contacts.[41]

IS further highlights the difference in dynamic behavior of mobile ions in Spiro 1 and Spiro 2 cells. Spiro 1 and Spiro 2 impedance responses are fit to a similar model as PTAA with slight modifications, as discussed below (Figs. 3d-f). Model fits over the full frequency range are shown in Supporting Figs. S4 and S5, and the extracted parameters are summarized in Supporting Fig.

S6. The bode plots for the representative PTAA, Spiro 1, and Spiro 2 cells are provided in Supporting Fig. S7. Overall, the recombination resistance, $R_1$, for the PTAA cell was lower than $R_1$ for either Spiro device, which correlates with the lower PCE, through lower $V_{OC}$ and FF, exhibited by the PTAA cells. The series resistance, $R_S$, measured in the PTAA device is about 1/3 that of the Spiro 1 device, which is consistent with the reduced thickness of the PTAA layer compared to the Spiro layer. In addition, the $R_S$ for the Spiro 2 device is greater than that of the Spiro 1 device, which correlates with their differences in performance. For Spiro 1 cells, at high bias (1 V), negative capacitance in the shape of a negative tail was observed in the impedance response, thus necessitating the use of an inductor element (L) in place of $C_2$ and a constant phase element with n = 0.75 in place of $C_3$ for all biases. For Spiro 2 cells, in the bias range between 0.6 V and 1.0 V, inductive loops are present in two distinct low frequency domains (Fig. 3e). These inductive loops are commonly observed for moderate PCE Spiro-OMeTAD cells.[6, 15, 47] In this case, the best fit was achieved using a modified equivalent circuit model[47] (Fig. 3f) with an inductor placed in parallel to the $R_2$/$C_2$ element (Supporting Fig. S5). These inductive loops are associated with the delayed mobile ion response due to surface polarization,[15] and thus corroborate our earlier suggestion that Spiro 2 ions are less mobile due to polarization from bias stressing.[48, 49] Equivalent circuit models for the Spiro 2 cells used constant phase elements in place of $C_3$. For low biases, the exponent, *n*, ranged between 0.7 and 0.8. On the other hand, for higher biases between 0.6 V and 1.0 V, *n* was 0.5.

We further consider the possibility that trap states in the bulk of the perovskite layer are a contributing factor to the observed LFN and IS data. Previously, the excess capacitance $C_S$ at low frequency has been correlated with the trap density of states *g(w)* within the perovskite layer, particularly for the case of a fully depleted absorbing layer, which is a reasonable assumption at low biases due to the large $V_{bi}$ and low intrinsic doping of $CH_3NH_3PbI_3$.[50] Under this assumption, the trap density $g(w)$ can be estimated from the slope of the capacitance $C(w)$ versus logarithm of frequency ($w$) in Fig. 2e by $g(w) = -\frac{V_{bi}}{qLk_BT}\frac{dC(w)}{d(\ln(w))}$, where q, $k_B$, T, L are the electronic charge, Boltzmann constant, temperature, and thickness of the perovskite layer, respectively.[14] By integrating $g(w)$ over frequency, one can infer that $C_S$ should scale with L. However, all PTAA, Spiro 1, and Spiro 2 cells have the same thickness (~400 nm) of the perovskite layer, while the $C_S$ (and $g(w)$ obtained from the slope of $C(w)$ versus $\ln(w)$) is roughly 2.5 times larger for PTAA than Spiro 1 or Spiro 2 (Fig. 2e). Thus, PTAA cells show 2.5 times larger trap density than Spiro cells. Consequently, the difference in $C_S$ and $g(w)$ cannot be reconciled by the bulk trap states. This analysis further supports our conclusion that the dominant LFN fluctuators result from an interfacial process (i.e., ion accumulation).

**Results and Conclusion**

For accumulated ions near the ETL, we obtain an effective Debye length ($L_D$) from $C_S = \varepsilon\varepsilon_0/L_D$, where $\varepsilon_0$ and $\varepsilon$ are the vacuum permittivity and dielectric constant of the perovskite, respectively. Taking $\varepsilon$ = 32.5 from literature precedent,[51] we obtain $L_D$ ~13 nm and ~2.8 nm for Spiro and PTAA cells, respectively, which implies a roughly 20 times larger density of accumulated charges (*n*) from $n = \varepsilon\varepsilon_0 k_B T / q^2 L_D^2$ in PTAA compared to both Spiro cells (*n* ~ 5.9 × 10$^{18}$ cm$^{-3}$ for PTAA, and *n* ~ 2.8 × 10$^{17}$ cm$^{-3}$ for Spiro cells).[14] In the ideal case, this excess accumulated charge does not contribute to additional $V_{OC}$ loss since $\Delta V_{OC}$ remains equal to the thermal energy from $\Delta V_{OC} = qnL_D/C_S = k_BT/q$.[14] However, recent experimental results have shown a $V_{OC}$ loss of ~80 mV for PSCs with the highest $V_{OC}$ of 1.26 V compared to the radiative limit of 1.34 V.[12, 52-54] So, the observed PCE and $V_{OC}$ losses in PTAA and Spiro 2, compared to Spiro 1, originate from either increased recombination at the interfaces or other degradation mechanisms such as trapped charges and chemical reactions that are beyond the simple model considered here.[13] Furthermore, current fluctuations could arise not only from the fluctuation of the number of carriers and mobility, but also from fluctuations in the energy barriers at interfaces.[17] A smaller $L_D$ would make energy barrier fluctuations more sensitive to the ions in PTAA cells compared to both kinds of Spiro cells. On the other hand, a relatively higher concentration of mobile ions in Spiro 1 would respond more effectively to AC signals at small frequencies, resulting in a cyclostationary process and low noise amplitude. This cyclostationary peak is only observable in the case of low noise cells and thus is only observed in Spiro 2 cells at low biases (< 0.2 V) (Fig. 3b).

The normalized noise power spectral density for all solar cells was found to have a stronger correlation with FF and PCE than $V_{OC}$ and $I_{SC}$ (Fig. 4 and Supporting Fig. S8). Noise amplitude is sometimes used as a predictor for solar cell performance metrics.[23] However, additional correlation with FF here suggests a common role of interfacial recombination in decreasing FF and increasing noise amplitude. Recently, processing advances have achieved high efficiency (>20%) and high quasi-Fermi level splitting using PTAA HTLs in solar cells.[55-57] We attribute the relatively low efficiency in our PTAA cells to the low doping level of the HTL. Finally, we note that although we focused on a conventional PSC geometry here, a previous 1/f noise study on inverted PSCs showed a similar transition from 1/f to 1/f$^3$ at ~200 Hz at 300 K.[20] Another study on impedance spectroscopy of inverted PSCs also showed an excess capacitance arising from the transition from dielectric to electrode polarization at ~200 Hz at 290 K.[58] Thus, we expect the conclusions drawn here to also be valid for inverted PSCs.

In conclusion, we have conducted correlated noise and impedance spectroscopy on a series of hybrid perovskite solar cells. All devices, using either Spiro-OMeTAD or PTAA HTLs, show current fluctuations with a characteristic response at 100 – 200 Hz that is related to electrode polarization from the accumulation of mobile ions near the $SnO_2$ ETL. Moderate PCE (~10%) in PTAA cells is accompanied by lower $V_{OC}$ due to losses at the HTL rather than the ETL. High PCE (~15%) Spiro-OMeTAD cells show low 1/f noise that is accompanied by a cyclostationary process around 200 Hz. Moderate PCE (~12%) Spiro-OMeTAD cells show three orders of magnitude larger 1/f noise than the high PCE devices, a Lorentzian feature, smaller capacitive but larger non-capacitive hysteresis in dark I-V curves, and large inductive loops in the impedance response. All of these observations are reconciled by permanent electrode polarization due to trapped ions near $SnO_2$. Finally, we show that normalized noise power spectral density for all solar cells correlates with FF and PCE. Overall, this study provides useful insights into low frequency carrier kinetics that can help inform ongoing efforts to reduce hysteresis and increase stability in field-deployable hybrid perovskite solar cells.

**Acknowledgements**

This research was supported by the Materials Research Science and Engineering Center (MRSEC) of Northwestern University (NSF DMR-1720139) and the Center for Light Energy Activated Redox Processes (LEAP), an Energy Frontier Research Center funded by the U. S. Department of Energy, Office of Science, Basic Energy Sciences under Award No. DE-SC0001059. Menghua Zhu gratefully acknowledges the Joint Educational PhD Program of the Chinese Scholarship Council (CSC). Kyle A Luck acknowledges a National Science Foundation

Graduate Research Fellowship. This work made use of the EPIC facility of the Northwestern University NUANCE Center, which has received support from the Soft and Hybrid Nanotechnology Experimental (SHyNE) Resource (NSF ECCS-1542205); the MRSEC program (NSF DMR-1720139) at the Materials Research Center; the International Institute for Nanotechnology (IIN); the Keck Foundation; and the State of Illinois.

**Experimental Methods**

*Material Preparation.* All reagents were analytical-grade and were purchased from Sigma-Aldrich, unless otherwise stated. Fluorine-doped tin oxide (FTO), with a sheet resistance of 14 Ω sq$^{-1}$ on glass, was purchased from Asahi Glass (Japan). The tin(II) oxide ($SnO_2$) ETL precursor solution was prepared by dissolving tin(II) chloride dehydrate ($SnCl_2·2H_2O$) in ethanol to a 0.1 M concentration.[29] The perovskite $CH_3NH_3PbI_3$ solution was prepared by dissolving 462 mg lead(II) iodide ($PbI_2$) and 159 mg methylammonium iodide (MAI) in 800 µL N,N-dimethylformamide (DMF) and 200 µL dimethyl sulfoxide (DMSO) followed by stirring at room temperature for 1 hr. The poly(triaryl amine) (PTAA, 99%, Sigma-Aldrich) HTL solution was prepared by dissolving 32 mg of PTAA and 3.6 mg of 4-isopropyl-4′-methyldiphenyliodonium tetrakis(pentafluorophenyl)borate (TCI America) in 1.6 ml of chlorobenzene. The 2,2',7,7'-Tetrakis[N,N-di(4-methoxyphenyl)amino]-9,9'-spirobifluorene (Spiro-OMeTAD, 99%, Sigma-Aldrich ) HTL solution was prepared by dissolving 72 mg of Spiro-OMeTAD in 1 mL of chlorobenzene, adding 6.8 μL 4-*tert*-butylpyridine (tBP) and 13.6 μL of a 28.3 mg mL$^{-1}$ solution of Li-bis(trifluoromethanesulfonyl) imide (Li-TFSI) in acetonitrile, and stirring at room temperature for 1 hr.

*Device Fabrication.* The cleaned FTO substrates with DI water and acetone were exposed to ultraviolet-ozone for 15 min. The SnO$_2$ precursor solution was spin-coated onto the treated FTO substrates at 2,000 rpm for 30 sec and thermally annealed at 180 °C in ambient for 1 hr. In a glove box under a nitrogen atmosphere, the perovskite solution was spin-coated onto the FTO/SnO$_2$ substrate at 500 rpm for 5 sec and 4000 rpm for 60 sec. After 10 sec of spin-coating at 4000 rpm, 600 μL of diethyl ether was drop-casted onto the substrate. The perovskite films were then annealed at 100 °C for 10 min in a nitrogen glove box. The PTAA solution was spin-coated onto the perovskite layer at 1500 rpm for 30 sec and thermally annealed at 70 °C for 5 min in the same glove box. The Spiro-OMeTAD solution was spin-coated onto the perovskite layer at 2000 rpm for 30 sec in a dry box without annealing. Lastly, an 80 nm thick gold (99.99% purity) layer was deposited by thermal evaporation. A metal mask was used to fix the active area of each solar cell at 0.09 cm$^2$.

*Device Characterization.* X-ray diffraction (XRD) spectra were measured using a Rigaku SmartLab X-ray diffractometer. SEM images were collected using a Hitachi SU8030 scanning electron microscope. Current-voltage measurements were performed in ambient under AM 1.5G light (100 mW cm$^{-2}$) using a solar simulator equipped with a Spectra-Nova Class A xenon arc lamp and calibrated with a silicon reference cell certified by NREL. Further electrical characterization occurred in a LakeShore CRX 4 K probe station, which has a base pressure of ~2 × 10$^{-5}$ Torr. Low-frequency noise measurements were conducted using a 1212 DL Instruments low noise current preamplifier and a Stanford Research Systems SR760 spectrum analyzer. During low-frequency noise measurements, a Keithley 2400 source-meter was used to bias the devices and measure device current. Impedance spectroscopy measurements were conducted in the dark using a Solartron Model 1260A impedance/gain phase analyzer. Measurements were collected for a forward bias range of 0 V to 1 V, a frequency range of 1 Hz and 1 MHz, and a 50 mV AC perturbation. The IS data was analyzed using licensed Z-View software (Scribner Associates Ltd.).

**Associated Content**

Supporting Information is available free of charge on the ACS Publications website and includes additional current voltage characteristics, histograms of device performance metrics, 1/f noise data, impedance spectroscopy data, and model fitting parameters.

**Figures**

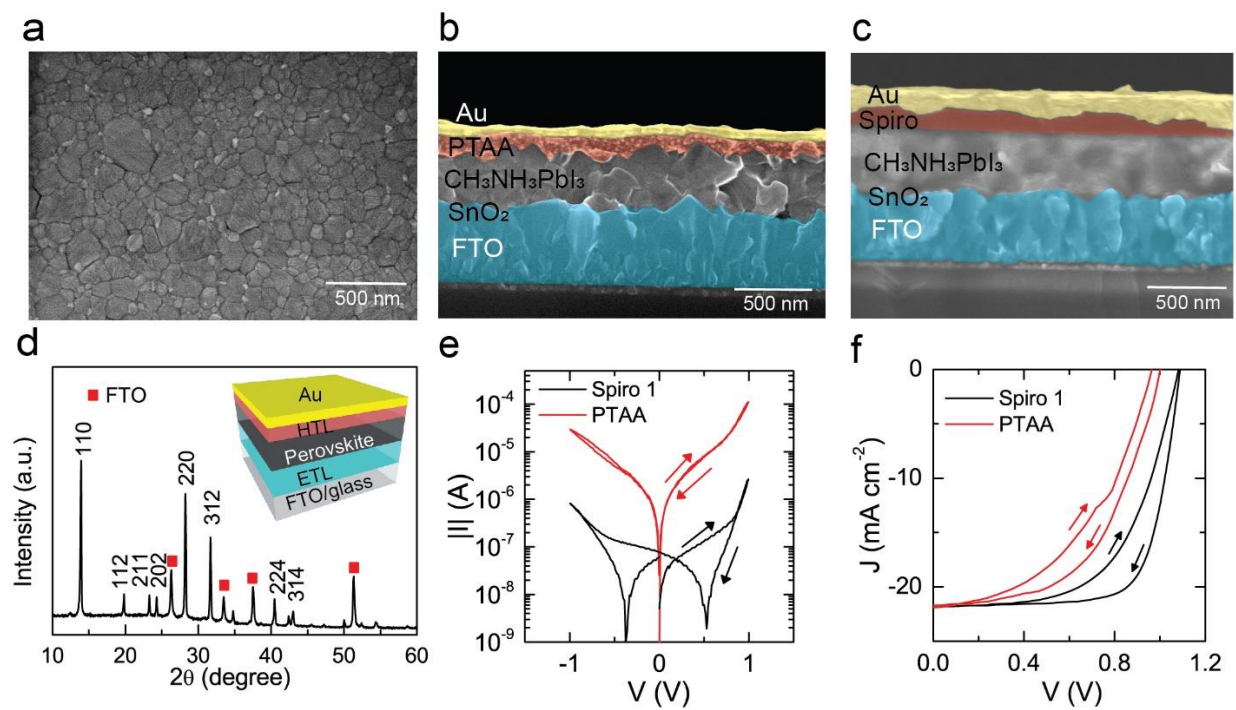

**Figure 1**. (a) Top-view scanning electron microscope (SEM) image of a $CH_3NH_3PbI_3$ perovskite film coated on a $SnO_2$/FTO substrate before deposition of the hole transfer layer (HTL). (b) Cross-sectional SEM images of a perovskite solar cell using PTAA as the HTL and $SnO_2$ as the electron transfer layer (ETL). (c) Cross-sectional SEM images of a perovskite solar cell using Spiro as the HTL and $SnO_2$ as the ETL. (d) X-ray diffraction pattern of the perovskite film showing crystalline perovskite peaks. FTO peaks are labeled with red squares. The inset shows a schematic of a perovskite solar cell (PSC). (e) Log-linear current-voltage (I-V) characteristics of PTAA and Spiro 1 PSCs in dark under vacuum (pressure $\sim 10^{-5}$ Torr). Arrows indicate the bias sweep direction. (f) Current-density-voltage (J-V) plot of PTAA (PCE $\sim$10%) and Spiro 1 (PCE $\sim$15%) PSCs under AM 1.5G light in ambient conditions. Dark I-V and illuminated J-V plots of Spiro 2 (PCE $\sim$12 %) are shown in Supporting Fig S1.

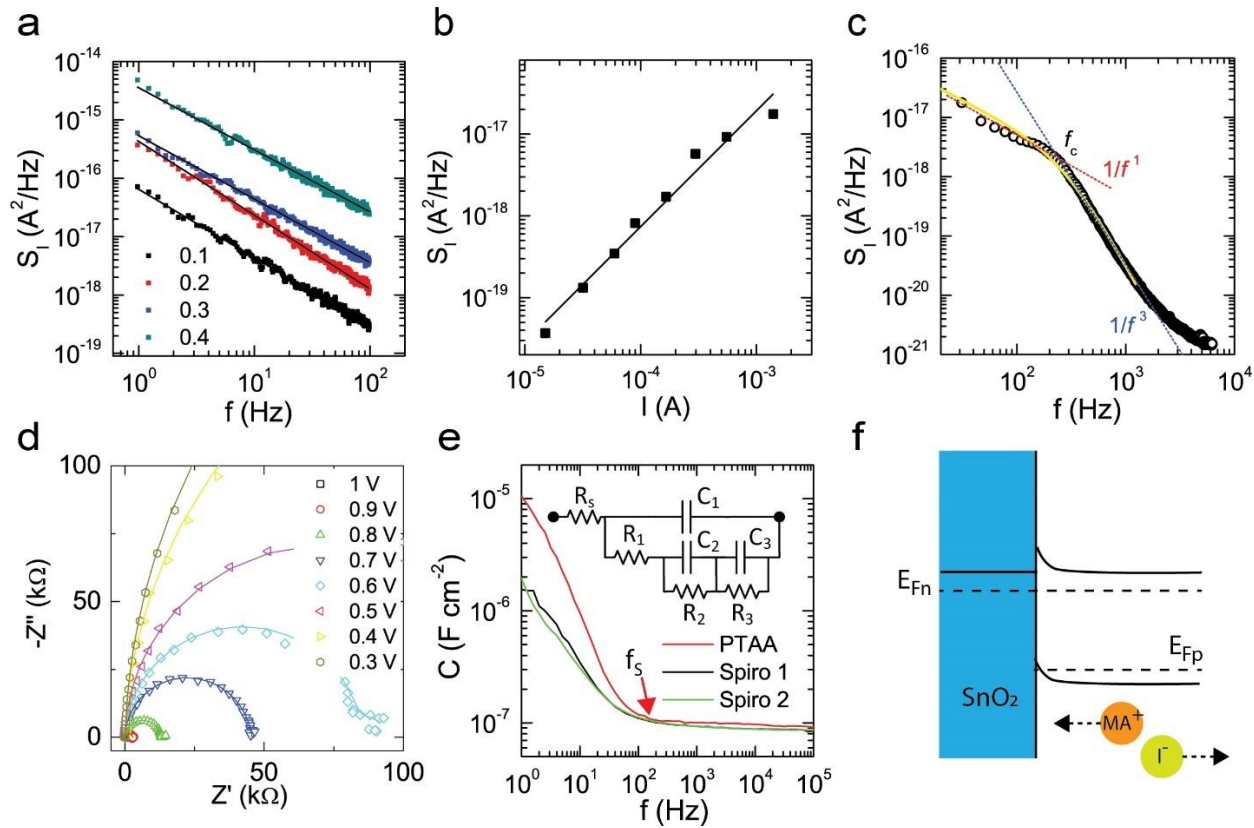

**Figure 2 | Noise and impedance spectroscopy for PTAA PSCs**. (a) Noise power spectral density (PSD) of a PTAA PSC at different biases under vacuum. Black lines are fits to the equation $S_I \sim 1/f^\beta$ with $\beta = 1.06 - 1.26$. (b) Noise PSD versus current at 1.95 Hz is fit to $S_I \sim I^\gamma$ with $\gamma = 1.49 \pm 0.1$. (c) Noise PSD showing transition from $1/f^{0.85}$ to $1/f^{3.0}$ behavior at a characteristic frequency ($f_c$) of 180 Hz. Yellow curve is a fit to equation (1). (d) Representative impedance response and model fits for a PTAA PSC at different biases under vacuum. All fit parameters are listed in Supporting Fig. S6. (e) Capacitance versus frequency of PTAA, Spiro 1, and Spiro 2 cells at 0.3 V, showing the transition between electrode polarization and dielectric polarization at a frequency $f_S$. The inset shows the equivalent circuit model used for fitting the data in (d). (f) Schematic showing perovskite band bending near the $SnO_2$ ETL in non-equilibrium conditions under

illumination or under DC biases in the dark, which drives methylammonium cations towards the ETL and iodine ions towards the HTL.

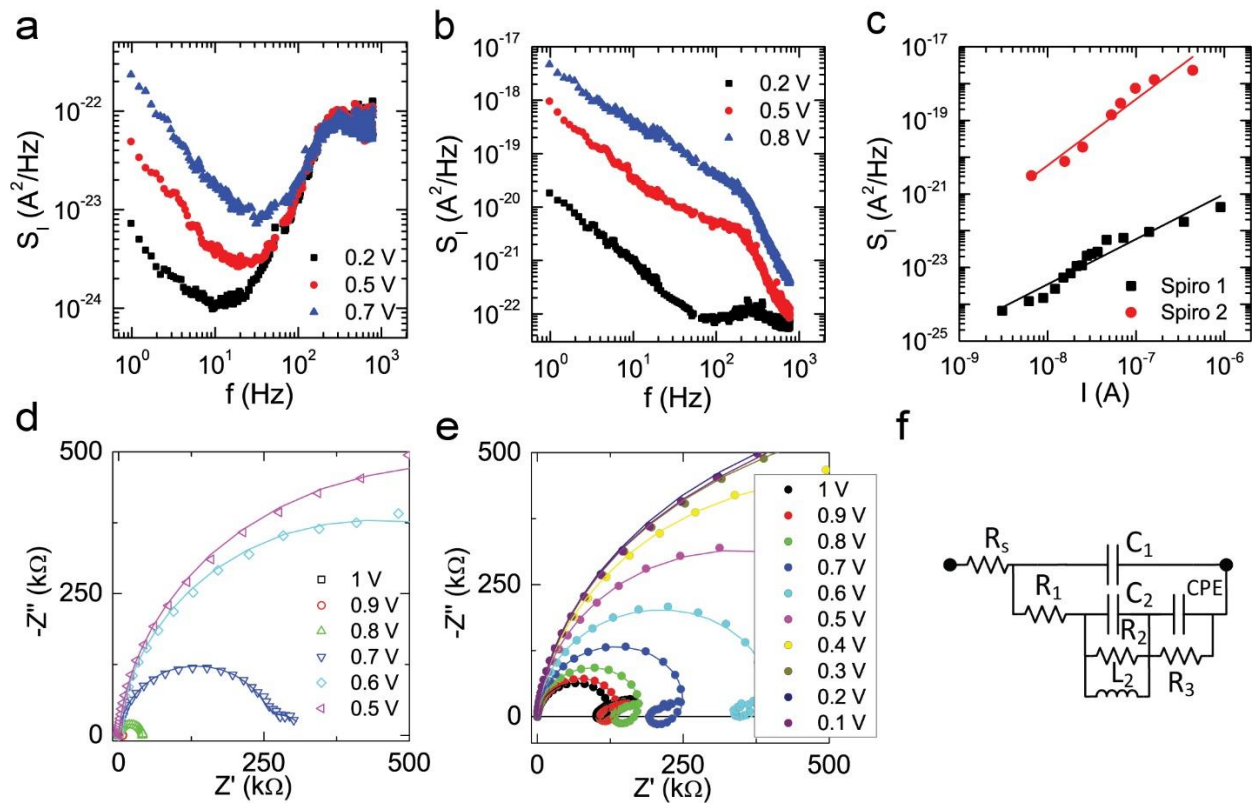

**Figure 3 | Noise and impedance spectroscopy for Spiro PSCs**. (a) Noise PSD of a Spiro 1 cell at selected values of applied biases under vacuum. (b) Noise PSD of a Spiro 2 cell at different biases. (c) Noise PSD versus current at 1.95 Hz is fit to $S_I \sim I^\gamma$ with $\gamma = 1.22$ and 1.79 for Spiro 1 and Spiro 2 cells, respectively. (d) Representative impedance response and model fits for a Spiro 1 cell under vacuum. (e) Representative impedance response and model fits for a Spiro 2 cell. (f) Schematic of the equivalent circuit model used for fitting the Spiro 2 cell. All fit parameters are listed in Supporting Fig. S6.

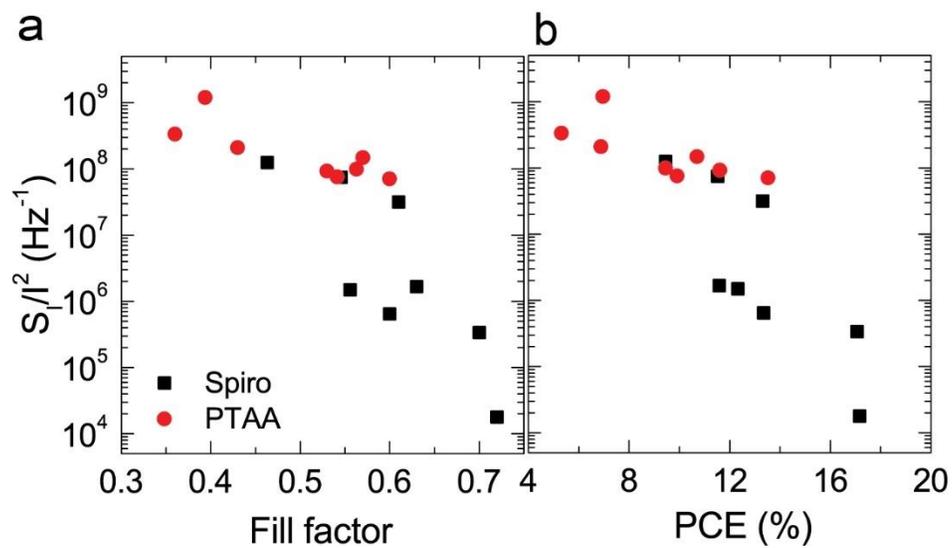

**Figure 4 | Correlation between noise amplitude and performance metrics**. A plot showing current normalized power spectral density versus (a) fill factor and (b) power conversion efficiency (PCE) for Spiro and PTAA solar cells. The noise spectral density shows a weaker correlation with $V_{OC}$ and $I_{SC}$ (see Supporting Fig. S8).

**TOC Image**

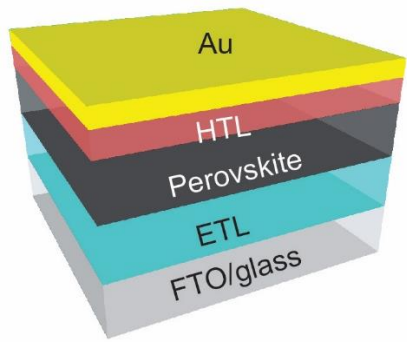 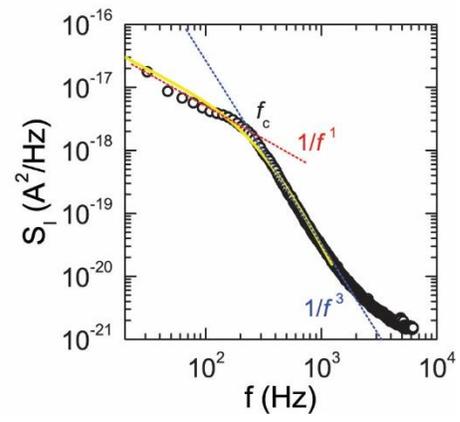